\documentclass[apj]{emulateapj}
\usepackage{graphicx, graphics}
\usepackage{natbib}
\received{}
\revised{}
\accepted{}

\shorttitle{}
\shortauthors{}

\begin{document}

\title{An HST/COS Observation of Broad Ly$\alpha$ Emission and Associated Absorption Lines of the BL Lacertae Object H~2356-309}

\author{Taotao~Fang\altaffilmark{1}, Charles W. Danforth\altaffilmark{2}, David A. Buote\altaffilmark{3}, John T. Stocke\altaffilmark{2}, J.~Michael Shull\altaffilmark{2}, Claude R. Canizares\altaffilmark{4}, and Fabio Gastaldello\altaffilmark{5}}

\altaffiltext{1}{Department of Astronomy and Institute of Theoretical Physics and Astrophysics, Xiamen University, Xiamen, Fujian 361005, China; fangt@xmu.edu.cn}
\altaffiltext{2}{CASA, Department of Astrophysical and Planetary Sciences, University of Colorado, 389-UCB, Boulder, CO 80309, USA}
\altaffiltext{3}{Department of Physics \& Astronomy, 4129 Frederick Reines Hall, University of California, Irvine, CA 92697}
\altaffiltext{4}{Department of Physics and Kavli Institute for Astrophysics and Space Research, Massachusetts Institute of Technology, Cambridge, MA 02139, USA}
\altaffiltext{5}{IASF-Milano, INAF, via Bassini 15, Milan I-20133, Italy}

\begin{abstract}

Weak spectral features in BL Lacertae  objects (BL Lac) often provide a unique opportunity to probe the inner region of this rare type of active galactic nucleus. We present a {\sl Hubble Space Telescope}/Cosmic Origins Spectrograph observation of the BL Lac H~2356-309. A weak Ly$\alpha$ emission line was detected. This is the fourth detection of a weak Ly$\alpha$ emission feature in the ultraviolet (UV) band in the so-called ``high energy peaked BL Lacs", after Stocke et al. Assuming the line-emitting gas is located in the broad line region (BLR) and the ionizing source is the off-axis jet emission, we constrain the Lorentz factor ($\Gamma$) of the relativistic jet to be $\geq 8.1$ with a maximum viewing angle of 3.6$^\circ$. The derived $\Gamma$ is somewhat larger than previous measurements of $\Gamma \approx 3 - 5$, implying a covering factor of $\sim$ 3\% of the line-emitting gas. Alternatively, the BLR clouds could be optically thin, in which case we constrain the BLR warm gas to be $\sim 10^{-5}\rm\ M_{\odot}$. We also detected two \ion{H}{1} and one \ion{O}{6} absorption lines that are within $|\Delta v| < 150\rm\ km\ s^{-1}$ of the BL Lac object. The \ion{O}{6} and one of the \ion{H}{1} absorbers likely coexist due to their nearly identical velocities. We discuss several ionization models and find a photoionization model where the ionizing photon source is the BL Lac object can fit the observed ion column densities with reasonable physical parameters. This absorber can either be located in the interstellar medium of the host galaxy, or in the BLR. 

\end{abstract}


\section{Introduction}

BL Lacertae objects, or BL Lacs, characterized by their strong flux, rapid variability and polarization, are often thought to be active galactic nuclei (AGNs) with relativistic jets beaming toward the direction of the observer (see, e.g., \citealp{blandford1978}). Between X-ray and radio, the spectra of BL Lacs are typically dominated by synchrotron radiation, and can be roughly classified as high-energy peaked BL Lacs (HBLs, peaked in ultraviolet to X-ray; \citealp{padovani1995}) or low-energy peaked BL Lacs (LBLs, peaked in infrared or longer wavelength).

The exact physical reason behind the LBL/HBL classification is still under debate. The ``blazar sequence" model suggests a diagram in which the peak frequency of the synchrotron emission is related to the electron energy (see, e.g., \citealp{fossati1998, ghisellini1998,ghisellini2008}, but also see \citealp{padovani2007}). Simple geometry (e.g., \citealp{nieppola2008}) or selection effects \citep{collinge2005} may also explain the difference between the two classes of BL Lacs.

Spectra of BL Lac objects often show no or very weak emission features, which is interpreted as the result of a very strong continuum boosted by relativistic beaming. Observationally, weak, optical emission lines such as H$\alpha$ have been observed in LBLs (e.g., \citealp{corbett2000, farina2012}). These lines originate in the broad line region (BLR) and therefore offer a direct probe of the inner regions around the central black holes. Recent observations show that the line-emitting gas does not respond to the rapid variation of the BL Lac continuum, suggesting that the BLR clouds are likely exposed by the ionizing photons from the accretion disk, instead of the synchrotron emission from the jet (e.g., \citealp{corbett2000, farina2012}).

In the X-ray regime, during 1980s and 1990s, a number of broad absorption lines were reported during the observations of BL Lac objects (see, e.g., \citealp{canizares1984, urry1986, madejski1991, grandi1997, sambruna1997}). They were often interpreted as highly ionized oxygen in a high-speed outflow (up to $\sim 10,000\rm\ km\ s^{-1}$) intrinsic to the BL Lacs (e.g., \citealp{krolik1985}). However, such broad features largely disappeared when observed with high-resolution spectrometers on board {\sl Chandra} and {\sl XMM}-Newton (\citealp{blustin2004, perlman2005}), leading to the conclusions that the previous detections were affected by a number of factors, such as the spectral quality, calibration, as well as continuum modeling. Recently, \citet{fang2011} reported the detection of a transient \ion{O}{8} absorption line during {\sl Chandra} observations of the BL Lac H~2356-309, indicating that further studies may be necessary to identify similar features in the X-ray spectra of BL Lacs.

While LBLs often present weak emission lines (e.g., \citealp{corbett2000}), little is known about HBLs. For HBLs, their redshifts are often revealed by weak absorption features such as \ion{Ca}{2} and \ion{Mg}{1} that were detected in the optical spectra of the host galaxies. Recently, using the Cosmic Origin Spectrograph (COS) on board the {\sl Hubble Space Telescope} ({\sl HST}), \citet{stocke2011} reported the detection of weak Ly$\alpha$ emission lines for the first time in three HBLs: Mrk~421, Mrk~501, and PKS~2005-489. Using the measured Ly$\alpha$ emission line flux, they were able to constrain either the Doppler beaming effect ($\Gamma$) or the mass of the warm gas in the BLR.

In this paper, we report the detection of a weak \ion{H}{1} Ly$\alpha$ emission line during our {\sl HST}/COS observation of the BL Lac object H~2356-309. After \citet{stocke2011}, this is the fourth HBL that presents such an emission line. Furthermore, we also discover three associated absorption lines. Such a combination of emission/absorption features provides a rare opportunity to study the inner environment of the BL Lac objects. This paper is organized as follows. Section \S2 presents the description of our observation and data reduction. We discuss the emission and absorption lines and their implications in Section \S3. The last section provides a summary.

\begin{figure*}[t]
\center
\includegraphics[height=0.3\textheight,width=0.49\textwidth]{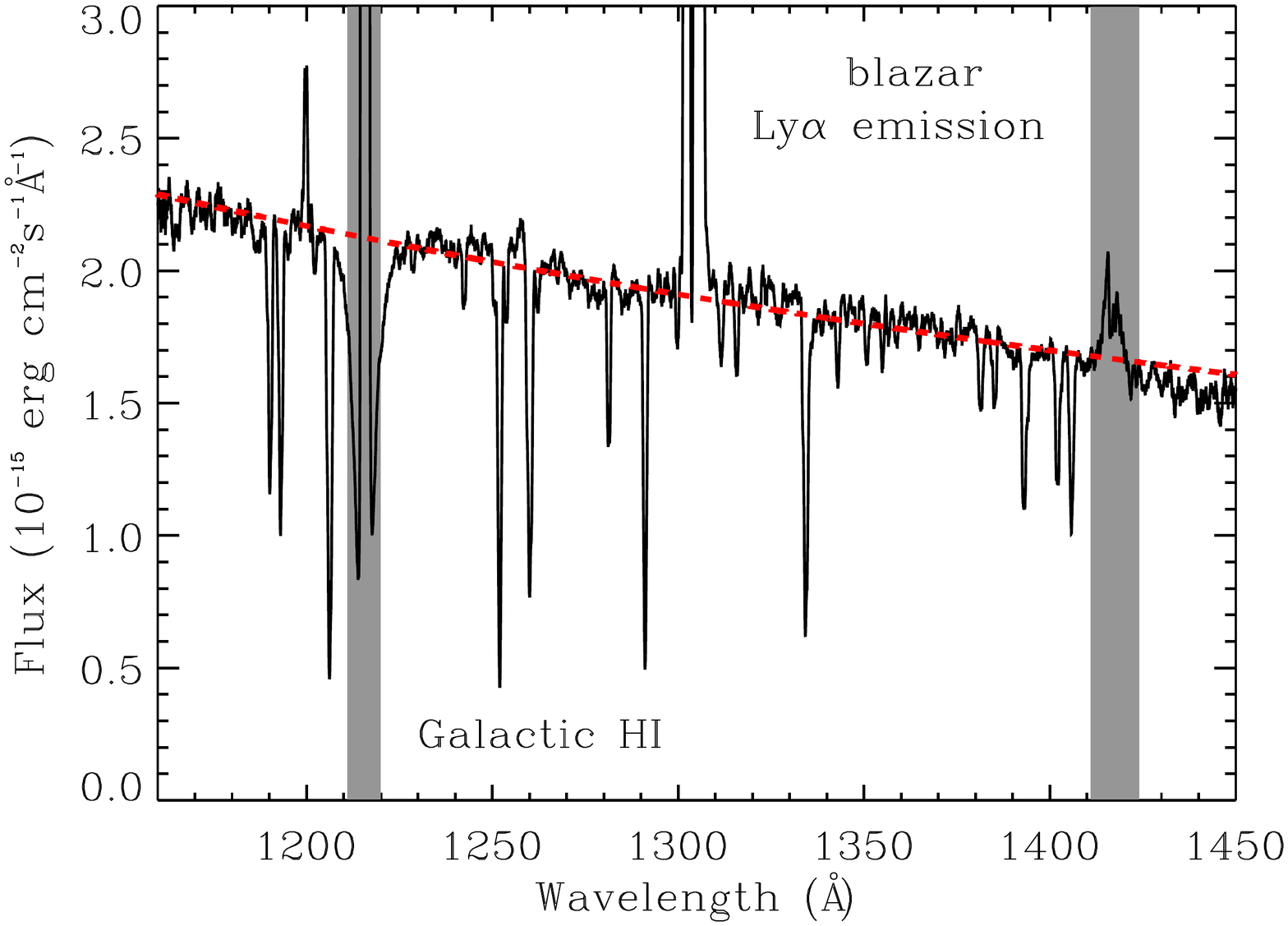}
\includegraphics[height=0.3\textheight,width=0.49\textwidth]{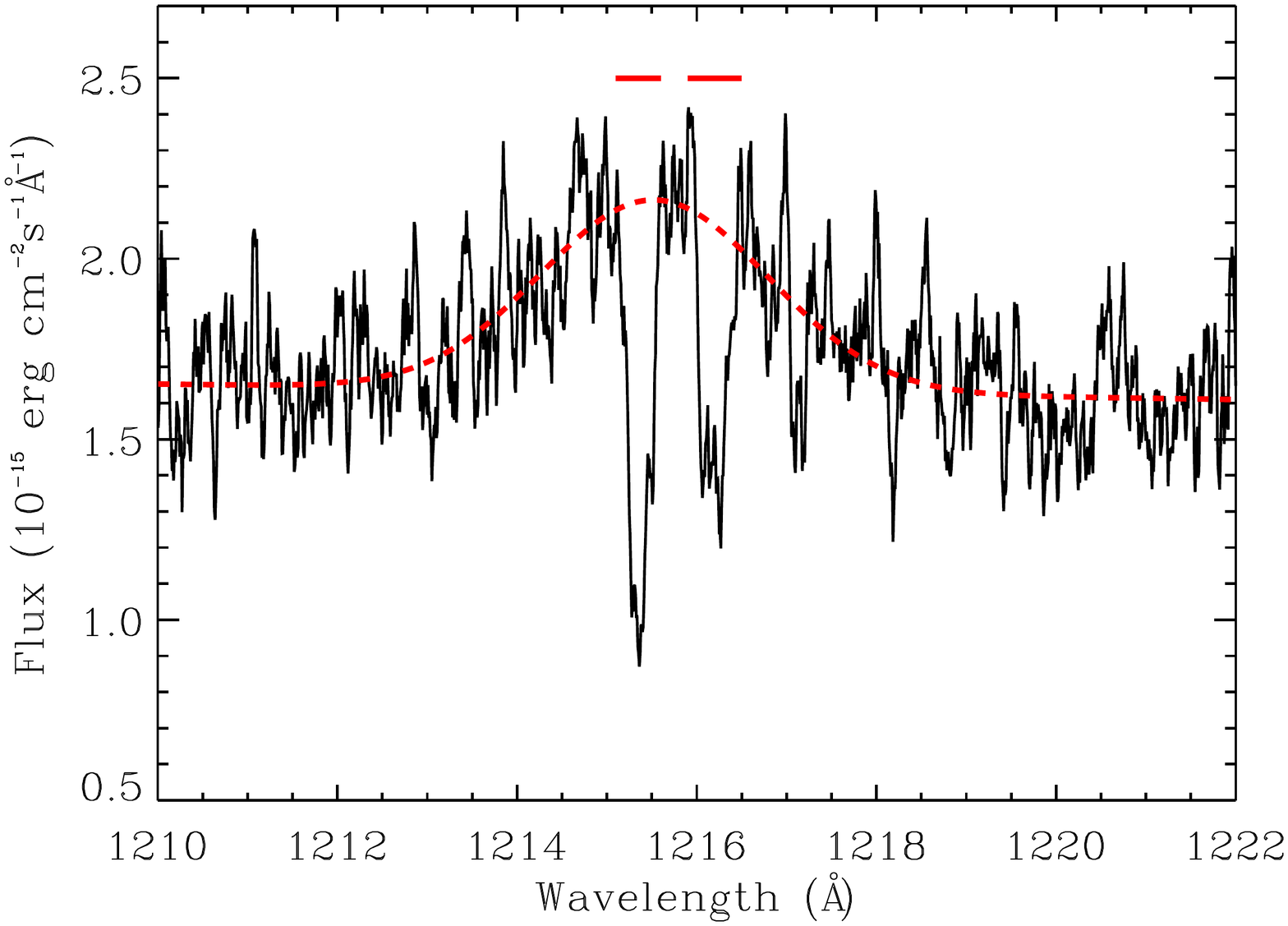}
\caption{Left panel: Broadband FUV spectrum of H~2356-309, between 1160 and 1450 \AA\ in the observer's frame. The two shadowed areas indicate the Galactic \ion{H}{1} Ly$\alpha$ absorption (left) and the blazar \ion{H}{1} Ly$\alpha$ emission (right). The red dashed line is a power-law fit to the continuum. Right panel: Blazar \ion{H}{1} Ly$\alpha$ emission line in the rest frame of the blazar. The red dashed line is a Gaussian fit to the emission line. The two horizontal red lines indicate the two \ion{H}{1} Ly$\alpha$ absorption features on the emission line profile. These horizontal lines also show the spectral regions that were cut when fitting the \ion{H}{1} Ly$\alpha$ emission.}
\label{fig:broad}
\end{figure*}

\section{Target and Observation}

\subsection{Redshift, Broadband Continuum, and Host Galaxy}

The most accurate measurement of the redshift of H~2356-309 ($\rm RA: 23^d59^m07^s.9$; $\rm Dec: -30^h37^m40^s.9$) was given by the 6dF Galaxy Redshift Survey (6dFGRS; \citealp{Jones2004, Jones2009}). The redshift, $z=0.16539\pm0.00018$ or $cz=49582\rm\ km\ s^{-1}$, was determined by cross-correlating the host galaxy spectrum with a set of predefined template spectra, following the procedure used for the 2dF Galaxy Redshift Survey \citep{colless2001}. The rms uncertainty is $47\rm\ km\ s^{-1}$. 

H~2356-309 is classified as an HBL because of its high X-ray flux \citep{giommi2005}. Multiband observations of H2356-309 from radio to very high energy (VHE, $E > 100$ GeV) suggested that the broadband spectral energy distribution (SED) can be simply fitted with a synchrotron self-Compton (SSC) model with a double-peak structure \citep{hess2010}, in which the peaks in X-ray and VHE are produced via synchrotron radiation and inverse Compton scattering, respectively. 

The host of H~2356-309 is a normal giant elliptical galaxy with a profile that can be well-fitted with a de Vaucouleurs law \citep{falomo1991}. Interestingly, with {\sl HST}/WFPC2, a blue companion was discovered about 1.2" away from the host galaxy, or at a separation of 4.6 kpc assuming the same redshift \citep{falomo2000}. We discuss the possible association of the detected absorption lines with this blue companion in Section \S3.

\subsection{Observation and Date Reduction}

H\,2356$-$309 was observed with the Cosmic Origins Spectrograph ({\sl HST}/COS) on 2013, June 15, for 17.0 ksec with the medium-resolution ($\Delta v\sim\rm 18~km~s^{-1}$), far-UV grating G130M (1135 \AA\ $<\ \lambda\ <$ 1450 \AA)\ as {\sl HST} program 12864.  The main science goal of this program was to search for broad \ion{H}{1}, Ly$\alpha$ and highly ionized metal absorbers corresponding to foreground superstructures, and we will present findings in this regard in a separate paper.  The unexpected scientific result of the observations we present here is the broad Ly$\alpha$ emission feature intrinsic to the BL\,Lac object as well as a trio of narrow absorption features at approximately the same redshift.

The flux-calibrated, one-dimensional spectra for each exposure were obtained from the Mikulski Archive for Space Telescopes (MAST) and combined using the standard IDL procedures described in detail by \citet{danforth2010}.  The combined data show a continuum flux level of $\sim1.7\times10^{-15}\rm~erg~cm^{-2}~s^{-1}~\AA^{-1}$ and a median signal to noise ratio (S/N) of $\sim12$  per seven-pixel resolution element.

The absolute wavelength calibration is crucial to the measurement of emission and absorption features. We quantified the uncertainty in the wavelength calibration by comparing the low-ionization Galactic absorption features detected in our COS spectrum with the \ion{H}{1} 21cm emission data. In the direction of H~2356-309, \citet{kalberla2005} shows a nice \ion{H}{1} profile with a velocity relative to the local standard of rest (LSR) of $v_{\rm LSR} = -5\rm\ km\ s^{-1}$. We then measured the centroids of as many low-ionized Galactic absorption lines (most are \ion{Si}{2}, \ion{Si}{3} and \ion{C}{2} lines) as we can find in the COS data. We found that the observed centroids are consistent with the LSR frame with an uncertainty of $\sim 10 \rm\ km\ s^{-1}$.

\section{Spectral Analysis}

\subsection{Continuum}

To accurately determine the far-UV continuum shape, Galactic extinction must be taken into account.  We normalized the data in the range 1180 \AA\ $<\ \lambda\ <$ 1250 \AA\ with a linear continuum, masked out regions of narrow emission (geocoronal airglow) and absorption (Galactic and extragalactic absorption features) and fitted a Voigt profile to the Galactic damped Ly$\alpha$ line (see Figure~\ref{fig:broad} left panel).  The resulting column density $\log N_{\rm HI} (\rm cm^{-2})=19.94\pm0.03$\footnote{All the errors are quoted at the 1$\sigma$ significance level unless otherwise stated.} implies an extinction of $E(B-V)=N_{\rm HI}/5.8\times10^{21}~{\rm cm^{-2}}=0.015\pm0.01$ via \citet{shull1985}. We assume no reddening in the AGN host galaxy.  We use this extinction to deredden the observed spectrum by applying the IDL routine {\sc fm\_unred}.

We then redshifted the spectrum into the rest-frame of the blazar, and fitted the broadband, dereddened continuum with a power- aw, $F_{\lambda} = F_{912}\left(\lambda/912{\rm\ \AA}\right)^{-\alpha_{\lambda}}$. To accurately characterize the continuum, we manually identified all the regions with strong absorption/emission features, and excluded those regions in our fit. We also smoothed the spectrum with a bin size of seven pixels. We find a power-law index of $\alpha_{\lambda} = 1.58\pm0.30$, and a normalization of $F_{912} = (2.63\pm0.16)\times\rm 10^{-15} erg\ cm^{-2}s^{-1}$. Using {\sl HST}/COS \citet{shull2012a} studied 22 AGNs and found a photon index of $\left<\alpha_{\lambda}\right>=1.32\pm0.14$ for the far UV composite spectra. This is consistent with what we found in H~2356-309. Figure~\ref{fig:broad} shows the spectrum between 1160 \AA\ and 1450 \AA\, in the observer's frame. The red dashed line is the continuum model. The two shadowed areas indicate the Galactic \ion{H}{1} Ly$\alpha$ absorption (left) and the blazar \ion{H}{1} Ly$\alpha$ emission which will be discussed later. While the overall fit is reasonably good, we clearly see deviations at some wavelength regions, particularly at around 1450 \AA.\ Such deviations are mainly caused by COS calibration uncertainties. The presence of fixed pattern noise can lead to 5 -- 10\% uncertainties in the absolute flux calibration\footnote{See http://www.stsci.edu/hst/cos/documents.}. Also, the detector edges are less sensitive and less well-calibrated than the middle of the detector since the spectrum is smeared out over a larger region of the detector. Such deviation will not affect our estimation of the ionizing photons at $\lambda<912$ \AA.

\subsection{Emission Line}

Due to the deviation of a single power-law fit to the broadband continuum at long wavelengths, we limited the fit to the range of 1210 -- 1222 \AA\ (in the rest frame of the blazar) when measuring the Ly$\alpha$ emission line flux. We also ignored the two \ion{H}{1} absorption line regions at $\sim$ 1215.5 \AA\ and 1217 \AA\ (the spectral regions labeled with horizontal red lines in the right panel of Figure~\ref{fig:broad}). We fitted the line in the rest-frame of the BL Lac and found that a single Gaussian profile fits the emission line very well (see the right panel of Figure~\ref{fig:broad}). We found the line central wavelength is $\lambda_{obs}=1215.53\pm0.14$ \AA,\ corresponding to $cz=49541\pm40\rm\ km\ s^{-1}$. This is consistent with the velocity of the host galaxy within 1$\sigma$ uncertainty. We also found the full width at half maximum (FWHM) is $5.43\pm1.30$ \AA,\ or $1340\pm321\rm\ km\ s^{-1}$, and the equivalent width is $EW=-1.03\pm0.22$ \AA. \ These values are slightly higher than those measured in \citet{stocke2011}, but at a substantially greater distance. We also derived an Ly$\alpha$ line intensity of $(1.68\pm0.36) \times 10^{-15}\rm\ erg\ cm^{-2}\ s^{-1}$ and a luminosity of $(9.53\pm2.02) \times \rm 10^{40}\rm\ erg\ s^{-1}$.

Following \citet{stocke2011}, we investigate the implication of the blazar Ly$\alpha$ emission line on our understanding of the BLR around the nucleus. We begin by calculating the expected Ly$\alpha$ flux under the ``nebular hypothesis". We assume the ionizing flux is isotropic, and the BLR clouds are optically thick to the ionizing photons. Using case-B recombination, the emission lines such as H$\beta$ can be estimated from the total number of the ionizing photons; Ly$\alpha$ luminosity can be calculated by the ratio of the specific intensity between Ly$\alpha$ and H$\beta$ (see \citealp{stocke2011} for detailed equations). We extrapolate the fitted continuum beyond the Lyman limit and estimate an Ly$\alpha$ emission line flux of $2.03 \times10^{-11}\rm\ erg\ cm^{-2}\ s^{-1}$, four orders of magnitude higher (over-prediction factor, or $OPF=1.2\times10^4$) than what we detect in our COS observation.

One way to explain the overprediction of the Ly$\alpha$ flux is simply that the ionizing flux seen by the BLR clouds is much weaker than what we see, due to the beaming effect of the blazar. The BLR clouds are likely exposed to the emission from either the accretion disk or the jet, or a combination of both \citep{yuan2014}. If the emission from the jet dominates, then we can estimate the required Lorentz factor ($\Gamma=1/\left(1-\beta^2\right)^{-1/2}$) and Doppler factor ($\delta=\left[\Gamma(1-\beta \cos \theta)\right]^{-1}$). Here, $\beta=v_j/c$ and $\theta$ is the viewing angle. The $OPF$ can be related to $\delta$ as: $OPF=\delta^{5-\alpha_{\lambda}}$ \citep{urry1995}. Using the power-law index from our continuum fitting, we find $\delta=16.1$. Since $\Gamma \geq (\delta+1/\delta)/2$ and $\sin\theta\leq1/\delta$, we obtain a minimum $\Gamma$ of 8.1 and a maximum viewing angle of 3.6$^{\circ}$ \citep{urry1995}. As observed in \citet{stocke2011}, if the true $\Gamma$ is typically as suggested $\Gamma=3-5$, our somewhat large value would imply a covering factor of $\sim$ 3\%. 

On the other hand, if the BLR clouds are mainly ionized by the flux from the accretion disk, then the derived Lorentz factor is likely to be a lower limit since the jet emission seen by the BLR must be much weaker \citep{stocke2011}. Detailed modeling may help separate the relative contributions from the disk and jet emission (see, e.g., \citealp{yuan2005, allen2006}).

A second, more likely interpretation is that instead of a weak ionizing flux, the weakness of the \ion{H}{1} emission line rather reflects a lack of line-emitting gas, either because the a small covering fraction of the BLR clouds, or because the BLR clouds are optically thin. Similar to the BL Lac objects studied in \citet{stocke2011}, the warm gas in the BLR of H~2356-309 would have a mass around $10^{-4}$ -- $10^{-5}\rm\ M_{\odot}$. A weak BLR implies a weak accretion disk and a weak accretion process with low radiative efficiency, such as those models suggested by the advection-dominated accretion flow (ADAF, see \citealp{yuan2014} for a review).

Finally, the broad line width suggests that the host galaxy of H~2356-309 is unlikely responsible for the observed Ly$\alpha$ emission line. A similar broad Ly$\alpha$ emission line was detected in an {\sl HST}/Faint Object Spectrograph observation of M~87 \citep{sankrit1999}. By comparing with a previous off-nucleus observation, \citet{sankrit1999} concluded that the majority of the Ly$\alpha$ emission comes from the central region of M~87. While several recent observations have suggested that giant ellipticals may not be as ``dead" as one expects and may host a large amount of cold gas (see, e.g., \citealp{thom2012, werner2014}), the detected H$\alpha$ line widths are significantly lower \citep{werner2014} than the Ly$\alpha$ line discussed in this paper. 

\begin{figure}[t]
\center
\includegraphics[height=0.35\textheight,width=0.5\textwidth]{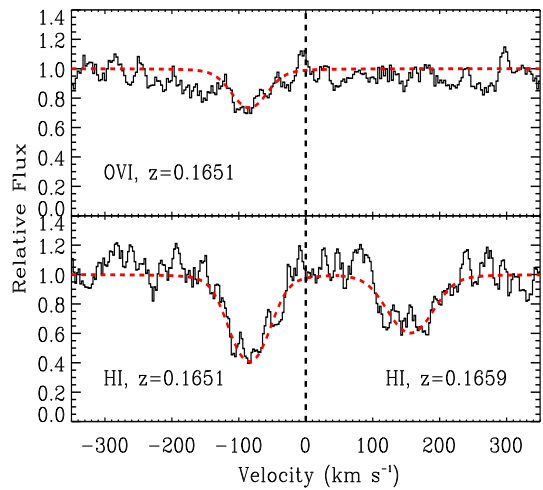}
\caption{Associated absorbers, \ion{O}{6} 1031.93 \AA\ (top) and \ion{H}{1} Ly$\alpha$ (bottom), are plotted in velocity space. The velocity of the host galaxy is defined as the zero-point (the vertical dashed line).}
\label{fig:abs}
\end{figure}

\subsection{Absorption Lines}

A number of Galactic and intergalactic absorption lines were identified in the {\sl HST}/COS spectrum of H~2356-309 and analyzed by Fang et al. (in preparation). In this paper, we focus on the three absorption lines that were identified as absorbers intrinsic to the blazar. Two of them are \ion{H}{1} Ly$\alpha$ absorbers (see the right panel of Figure~\ref{fig:broad}). We also marginally detected the stronger line $\lambda$1031.93 \AA\ of the \ion{O}{6} doublet at the blazar redshift. 

The absorption lines were fitted with custom IDL routines using a Voigt-based line profile (see \citealp{danforth2011} and Keeney, private communication). The stronger \ion{H}{1} Ly$\alpha$ ($\rm EW=166\pm17$ m\AA)\ is located at $cz=49498\pm4\ \rm km\ s^{-1}$, well within the rms uncertainty of the blazar redshift (see the top panel of Figure~\ref{fig:abs}). Its profile is well fitted with $b=29\pm5\rm\ km\ s^{-1}$, and $\log N_{HI} (\rm cm^{-2})= 13.64\pm0.09$. To estimate the line significance level (expressed in Gaussian $\sigma$, we make use the empirical relation derived in \citet{keeney2012} and \citet{danforth2011} (their equation [4]): $\rm SL = 0.095 W_{\lambda} (S/N/pix) / b^{0.62} = 9.0$. Here $W_{\lambda}$ is the line EW, $\rm (S/N/pixel)$ is the S/N per COS pixel (see Table~1).

A weak absorption feature is detected at 1202.3 \AA.\ No commonly known Galactic interstellar medium (ISM) lines are located around this wavelength; the nearest \ion{N}{1} lines are at around 1200 \AA.\ Such feature also cannot be redshifted \ion{H}{1} Ly$\alpha$ or Ly$\beta$. This feature, however, if it belongs to an absorber intrinsic to the BL Lac object, fits well with the stronger transition of the \ion{O}{6} $\lambda\lambda$ 1031.93, 1037.67 doublet. We tentatively identify this feature as an \ion{O}{6} absorption line at the redshift of H~2356-309. Our Voigt profile shows a line EW of $58\pm14$ m\AA\, a column density of $\log N_{OVI} (\rm cm^{-2})=13.73\pm0.12$, and $cz=49497\pm8\rm\ km\ s^{-1}$. 

Giving the proximity of the $z=0.16509$ \ion{O}{6} and \ion{H}{1} absorption lines ($|\Delta v| = 3\rm\ km\ s^{-1}$), it is interesting to consider the possibility that a single-phase absorber is responsible for both absorption lines. In this case we can estimate the temperature of the absorber by assuming the broadening of both lines is a combination of thermal and non-thermal (such as turbulence) effects. We can write $b^2 = 0.129^2 (T/A) + b^2_{nt}$, where $b$ and $b_{nt}$ are the Doppler-$b$ parameter and the non-thermal velocity in units of $\rm km\ s^{-1}$, respectively. By solving this equation for the \ion{H}{1} and \ion{O}{6} lines, we can then find the temperature and the non-thermal velocity of the absorber. It is unlikely that the Doppler-$b$ parameters for both lines are at their best-fit values, since in this case $b(HI) < b(OVI)$ and the temperature would then be negative. However, we can estimate the upper limit of the temperature by adopting the the 1$\sigma$ upper limit for the Doppler-$b$ parameter of the \ion{H}{1} line and the 1$\sigma$ lower limit for the \ion{O}{6} line. We found a maximum temperature of $T= 7.7\times10^4$ K and a minimum non-thermal velocity of $b_{nt}=11.4\rm\ km\ s^{-1}$ at this temperature. 

\begin{deluxetable*}{lcclcccc}
\tablewidth{0pt}
\tiny
\tablecaption{Absorption Line List}
\tablehead{$z_{abs}$ & Line & $\lambda_{obs}$ & $W_r$ & $b$ & $\log N$ & $\rm S/N/pix$ & SL$^a$\\
 & & (\AA) & (m\AA) & ($\rm km\ s^{-1}$)& ($\rm N_{ion}, cm^{-2}$) & & }
\startdata

0.16508 & \ion{O}{6} $\lambda$ 1031.93 & 1202.30 & $58\pm14$& $26\pm11$ & $13.73\pm0.12$ & 6.3 & 4.6 \\
0.16509 & \ion{H}{1} Ly$\alpha$ & 1416.36 & $166\pm17$ & $29\pm5$ & $13.64\pm0.09$ & 4.6 & 9.0\\
0.16590 & \ion{H}{1} Ly$\alpha$ & 1417.34 & $124\pm18$ & $35\pm8$ & $13.44\pm0.14$ & 4.6 & 6.0
\enddata
\label{tab:abs}
\tablecomments{a. Significance level is estimated using the empirical relation derived in \citet{keeney2012} and \citet{danforth2011} (their equation [4]): $\rm SL = 0.095 W_{\lambda} (S/N/pix) / b^{0.62}$ (see text).}
\end{deluxetable*}

Such an absorber is unlikely to be in a pure collisional ionization equilibrium. The ratio between the measured column densities of the two species implied a temperature range of $10^{5.2}$ -- $10^{6.5}$ K, for metal abundances between 0.1 and 1 solar abundance. This temperature range is inconsistent with the upper limit we inferred from the Doppler-$b$ parameters of the \ion{H}{1} and \ion{O}{6} lines. As suggested by \citet{tripp2008}, the line ratio can also rule out the non-equilibrium, radiatively cooling process modeled by \citet{gnat2007}

We also investigate the physical condition of the absorber under the assumption of photoionization. Numerous \ion{H}{1} absorbers with associated \ion{O}{6} absorption were discovered in the past decade (see, e.g., \citealp{danforth2005,danforth2006,tripp2008}). These absorbers are distributed in the intergalactic medium (IGM) and are likely to be under the influence of the cosmic UV background. Adopting the \citet{haardt1996, haardt2012} UV background, we estimate physical properties of the absorber using the photoionization code CLOUDY (version 13.02, last described by \citealp{ferland2013}). The calculation stops when the \ion{H}{1} column density reaches $\log N_{HI} (\rm cm^{-2}) = 13.53$. We find that an ionization parameter of $\log \xi=-0.9$ for a metal abundance of $\log Z =-0.5$ is necessary to produce the observed \ion{O}{6} column density. Here, $Z$ is metallicity in units of solar abundance, and the ionization parameter is defined as $\xi  = \phi/n_H c$, where $\phi$ is the surface flux of ionizing photons, $n_H$ is the hydrogen density, and $c$ is the speed of light. However, the required $n_H$ is $10^{-5.5}\rm\ cm^{-3}$, suggesting a path length of 1.5 Mpc for the absorber. The Hubble broadening at this scale is $b_H \approx 50\rm\ km\ s^{-1}$ assuming a Hubble constant of $70\rm\ km\ s^{-1}Mpc^{-1}$ \citep{danforth2008}, and is larger than the observed line width. Therefore, we also rule out the photoionization model of the \ion{O}{6}-\ion{H}{1} absorber in the IGM, unless the metallicity is much higher than the typical value for a low-density IGM environment.

This absorber also cannot be associated with the ISM/halo gas in foreground galaxies. Using {\sl HST}/WFPC2, \citet{falomo2000} found a companion galaxy about 1" away from the sightline toward H~2356-309. A recent Magellan/IMACS survey also found several galaxies within 150" (or an impact distance of 400 kpc at the blazar redshift) of this sightline (\citealp{williams2013}; Williams, private communication). Using the same UV background we find even if we apply the solar metal abundance, the required pathlength would still be around 1 Mpc which is far too large for an absorber to be within a galaxy.   

The \ion{O}{6}-\ion{H}{1} absorber is $\sim100\rm\ km\ s^{-1}$ away from the blazar and can be classified as the so-called ``proximity absorber" (\citealp{weymann1979, foltz1986}). It is likely that the absorber is ionized by the UV radiation from the blazar. We consider two scenarios: the absorber is either located in the ISM of the host galaxy of H~2356-309 or in the BLR. We ran the CLOUDY calculation with the input spectrum derived in the previous section, and find the observed column densities can be achieved in both scenarios with reasonable physical parameters. If the absorber is located in the ISM, assuming a solar abundance and a hydrogen density of 1 $\rm cm^{-3}$, we find the absorber has an ionization parameter of $\log \xi=-1.2$, a size of 2.6 kpc, and a distance to the source of 2.7 kpc. For the BLR case, we also assume a solar abundance but with a hydrogen density of $10^{10}\rm\ cm^{-3}$. We find an ionization parameter of $\log \xi=-1.4$, a size of 0.026 pc, and a distance to the source of 0.034 pc.

The second \ion{H}{1} Ly$\alpha$ absorption line is located at the red side of the broad Ly$\alpha$ emission line, with a velocity of $cz=49738\pm7\ \rm km\ s^{-1}$. This is about $150\rm\ km^{-1}$ away from the emission line redshift. Such a phenomenon is frequently observed in the Ly$\alpha$ forest spectra of quasars (see, e.g., \citealp{scott2002}); however, this is the first time it was detected in the BL Lac spectrum. It normally suggests the influence of inflow or a large-scale peculiar velocity (see, e.g., \citealp{loeb1995}). The derived properties of this \ion{H}{1} absorber are similar to those of the \ion{H}{1} Ly$\alpha$ at $z=0.1651$ at a weaker significance level (see Table~\ref{tab:abs}). 

\section{Summary and Discussion}

In this paper, we present our {\sl HST}/COS observation of the BL Lac H~2356-309. We summarize our findings here.

\begin{itemize}

\item We observed the BL Lac H~2356-309 on 2013 June twice, with a total exposure of $\sim$ 17 ksec. We obtained a moderate spectrum with an S/N of $\sim 5$. The broadband, dereddened continuum $F_{\lambda}$ can be reasonably well fitted by a power law with a spectral index of $\alpha_{\lambda}=1.58\pm0.30$.

\item A weak \ion{H}{1} Ly$\alpha$ emission line was detected at the redshift of H~2356-309. We model this emission line by assuming the line-emitting gas in the BLR is ionized by the jet radiation. We find the Ly$\alpha$ emission is over predicted by a factor of $\sim 10^4$. One explanation is that the continuum we observed is Doppler boosted, and we estimate the minimum Lorentz factor $\Gamma$ and maximum viewing angle. Alternatively, it may reflect a lack of line-emitting gas in the BLR. We estimate the mass of the warm gas in the BLR. Such weak BLR in general suggests a weak accretion process with low radiative efficiency, such as ADAF.

\item We also detect three associated absorption lines. Two of them (\ion{O}{6} 1031.93 and \ion{H}{1} Ly$\alpha$) have well-aligned velocities, suggesting they are probably produced by a single-phase absorber. A pure collisional ionization model for this absorber is ruled out because the temperature inferred from the line ratio is much higher than that derived from the line width. We also rule out photoionization by the cosmic UV background because the Hubble broadening suggested by the path length of the absorber would be larger than the observed line width. However, photoionization by the BL Lac object can explain the observed column densities, whether the absorber is located in the ISM of the host galaxy or in the BLR.  

\end{itemize}

The host galaxy of H~2356-309 is a giant elliptical galaxy. Such early-type galaxies, typically described as``red and dead", lack gas and star formation activities and one may expect to detect little or no \ion{O}{6} or \ion{H}{1} absorbing gas in its ISM.  However, a recent survey by \citet{thom2012} detected strong \ion{H}{1} absorbers around these early-type galaxies, suggesting the presence of a significant amount of cool, photoionized gas. \citet{werner2014} also reported the detection of cold and warm gas in nearby X-ray bright elliptical galaxies, indicating that these early-type galaxies may not be as ``dead" as we expected.

Shock waves have been suggested as the main heating mechanism for the warm-hot phase IGM at low redshift (\citealp{cen1999,dave2001,kang2005}). Significant amounts of \ion{O}{6} can be produced in the radiatively cooling gas behind the shock, even when the postshock temperature is significantly different from the peak temperature required for the maximum \ion{O}{6} ionization fraction under collisional ionization equilibrium \citep{heckman2002}. Following \citet{stocke2006}, we consider the velocity separation between the two \ion{H}{1} absorbers, $(\Delta v)_{Ly\alpha} = 250\rm\ km\ s^{-1}$ to be the one-dimensional shock velocity. The postshock temperature is $T_s = (1.34 \times 10^5 {\rm K}) (V_s/100\rm\ km\ s^{-1})^2$. Here $V_s = \sqrt 3(\Delta v)_{Ly\alpha}$ is the shock velocity. We find a shock temperature of $T_s = 8.4\times10^5$ K. Again, this temperature is inconsistent with the upper limit derived from the line width, suggesting that the shock is not the main mechanism for metal production here.

Between 2007 and 2008, H~2356-309 was observed extensively in X-ray for the study of the warm-hot IGM in the foreground galactic superstructures (\citealp{buote2009, fang2010}). During one of the {\sl Chandra} observations, a transient \ion{O}{8} K$\alpha$ absorption line intrinsic to the blazar was detected with high significance \citep{fang2011}. The temperature and ionization parameter for the X-ray absorber are significantly different from those of the UV absorbers, suggesting a multi-phase distribution of the warm to hot gas along this sightline. 

Since \citet{stocke2011}, this is the fourth BL Lac showing a weak \ion{H}{1} Ly$\alpha$ emission line in the UV band. Our ability to jointly observe both the continuum and the Ly$\alpha$ emission line provides a unique opportunity to probe the line-emitting region, presumably the BLR, of the BL Lac objects. Future observations are necessary to understand the nature of these emission/absorption features. For example, similar to the optical observations, monitoring these BL Lac objects and examining whether the emission line responses to the continuum variation can help explain the likely source of the ionizing photons. 

\acknowledgments
We appreciate helpful discussions with Feng Yuan, Tinggui Wang and Rik Williams. T.F. was partially supported by the National Natural Science Foundation of China under grant No. 11273021, by the  ``Fundamental Research Funds for the Central Universities" No. 2013121008, and by the Strategic Priority Research Program "The Emergence of Cosmological Structures" of the Chinese Academy of Sciences, Grant No. XDB09000000. J.M.S. thanks the Institute of Astronomy, Cambridge University, for their support through the Sackler visitor program.


\end{document}